\documentclass{INTERSPEECH2023}

\interspeechcameraready

\usepackage[natbib, bibencoding=utf8, citestyle=numeric, bibstyle=ieee, maxbibnames=999, maxcitenames=2, mincitenames=1, sortcites]{biblatex}
\bibliography{mybib}

\usepackage[capitalise]{cleveref}
\usepackage{soul}
\usepackage{dblfloatfix}
\usepackage[activate]{microtype}
\title{Abusive Speech Detection in Indic Languages Using Acoustic Features}
\name{Anika A. Spiesberger$^1$, Andreas Triantafyllopoulos$^1$, Iosif Tsangko$^1$, Björn W. Schuller$^{1,2,3}$}

\address{
  $^1$Chair of Embedded Intelligence for Health Care and Wellbeing, University of Augsburg, Germany \\
  $^2$audEERING GmbH, Gilching, Germany \\
  $^3$GLAM – Group on Language, Audio, \& Music, Imperial College, UK}
\email{anika.spiesberger@uni-a.de}

\begin{document}

\maketitle

\begin{abstract}
Abusive content in online social networks is a well-known problem that can cause serious psychological harm and incite hatred. The ability to upload audio data increases the importance of developing methods to detect abusive content in speech recordings. However, simply transferring the mechanisms from written abuse detection would ignore relevant information such as emotion and tone. In addition, many current algorithms require training in the specific language for which they are being used. This paper proposes to use acoustic and prosodic features to classify abusive content. We used the ADIMA data set, which contains recordings from ten Indic languages, and trained different models in multilingual and cross-lingual settings. Our results show that it is possible to classify abusive and non-abusive content using only acoustic and prosodic features. The most important and influential features are discussed.
\end{abstract}

\section{Introduction}
\label{sec:introduction}
With the rise of the internet and the establishment of more and more online social networking platforms, the amount of abusive content is rapidly increasing. The fact that almost every online social network provides information on how to prevent bullying, how to deal with abusive behaviour, and even statistics on the prevalence of harassment on its website underlines the growing importance of this issue. According to \citet[p.~274]{kaur2021abusive}, abusive behaviour ``describes a broad category of content that includes hate speech, profanity, threats, cyberbully and various ethnic and racial slurs''. It is often conveyed through rhetorical devices such as aggressive language, threats, obscenity, and sarcasm~\citep{poletto2021resources}.

In addition to hindering social interaction and fomenting hatred, research has shown that victims of cyberbullying ``have psychosomatic problems \ldots, have high levels of perceived difficulties, have emotional and peer problems, and feel unsafe at school and uncared about by teachers''~\citep[p.~727]{sourander2010psychosocial}. These findings increase the need to identify abusive content in online social networks. However, \citet{matamoros2021racism} found out that, to date, much of the research in this area has focused on the English language and on detecting hate speech in texts or memes. Their paper shows that most studies have been conducted in North America (44\%) and the United Kingdom (11\%) and have used data from Twitter (55\%) and Facebook (35\%). This imbalance likely exists because the majority of data sets come from these platforms.

As almost all social networking platforms now allow the upload of video and audio files, online communication and interactions are no longer limited to written content. In recent years, several efforts have therefore been made to classify abusive speech in audio recordings. Different approaches have been used for feature extraction and classification. For example, \citet{wu2020detection} used a speech-to-text converter and sentiment analysis to obtain the polarity of spoken utterances in audio recordings extracted from YouTube videos, \citet{ghosh2021detoxy}, who created DeToxy, one of the few available data sets of abusive audio recordings, used automatic speech recognition (ASR) for their baseline estimation, and \citet{ablaza2014suppression} used keyword spotting to suppress predefined profane words.  

While some information about abusive content in audio data can be retrieved using ASR and textual search or keyword spotting, this ignores the information provided by emotional and acoustic cues~\citep{yousefi2021audio}. These cues can be important, for example for phrases that may or may not be considered abusive depending on the tone and context. For this reason, the change of modality triggers the need for recognition methods that are specifically designed for audio and video data~\citep{kaur2021abusive, yang2022multimodal, vidgen2019challenges, alrashidi2022review}. 

\citet{sharon2022multilingual} combined the features extracted by ASR with contextual and emotional representation features. While they achieved F1 scores between 0.77 and 0.85 for the different languages in the ADIMA data set, the features they used contained only spectral and temporal information, and not prosodic or voice quality features, which have been shown to correlate with abusive speech. For example, \citet{novitasari2018rudewords} and \citet{sutejo2018indonesia} found that it is possible to detect rude words using only acoustic features. Although the performance of their textual model was better, this could be because they used acted data instead of real-life recordings, where the text was explicitly designed to be abusive. Furthermore, their findings might be relativised if the conversion of audio to text is complicated, resulting in less accurate textual representations. This may be the case, among other things, due to slurred speech and poor audio quality, such as background noise~\citep{yousefi2021audio}.

A popular method for obtaining acoustic features is using spectrogram images of the audio recordings. So far, this has been done mainly for Indic languages, once by \citet{rahut2020bengali}, who tried to distinguish between abusive and non-abusive content in Bengali speech recordings, and then by \citet{gupta2022adima} for baseline calculations on their newly introduced ADIMA data set, which contains abusive and non-abusive audio recordings for ten different Indic languages. \citet{gupta2022adima} additionally used Wav2Vec2 models for speech recognition. Both papers applied a variety of classification methods and achieved good overall results. In addition, \citet{gupta2022adima} found that performance in some languages was higher when the model was trained in another language. In particular, training in all languages improved performance in most single languages. If this could be generalised to other languages, it would be a major advantage for abusive content classification in audio data. However, while they hypothesise that ``models are able to leverage audio properties like pitch, emotions, intensity etc.\ for this task instead of relying on the actual word'' \cite[p.~6175]{gupta2022adima}, this could also be because the Wav2Vec2 models they used for speech recognition were pre-trained on the target languages. Therefore, as seen in \citep{Triantafyllopoulos22-PSE}, these are not true zero-shot trials and further investigation is needed, as large pre-trained models can exploit linguistic information, which may partially explain their higher performance.

This paper investigates how paralinguistic features alone, i.\,e., acoustic and prosodic properties, perform for the detection of abusive content in real-life data. This is particularly interesting because real-life recordings usually contain a lot of information about the speaker's emotional state that is not captured by the text. We also look at their performance in multilingual and cross-lingual settings, as non-textual features could provide an easy way to perform abusive content detection in less common languages for which no audio data sets are available.

The remainder of this paper is organised as follows: \cref{sec:methodology} describes the methodology used to distinguish which features differ significantly between abusive and non-abusive content, and how the classification was performed. \cref{sec:results} shows the results, which are followed by the discussion in \cref{sec:discussion} and the conclusion in \cref{sec:conclusion}.

\section{Methodology}
\label{sec:methodology}
The ADIMA data set~\citep{gupta2022adima} was used for this study. This data set consists of 11,775 audio recordings, evenly distributed across ten different Indic languages (Bengali, Bhojpuri, Gujarati, Haryanvi, Hindi, Kannada, Malayalam, Odia, Punjabi, and Tamil), and taken from real-life conversations in ShareChat chatrooms. The authors defined the content as abusive if swear words, cuss words, or abusive words/phrases were present, resulting in 5,108 abusive and 6,667 non-abusive recordings. The average duration of the recordings is 20~($\pm$ 3) seconds with a minimum of 5 seconds and a maximum of 58 seconds. For each language, the authors split the data set into a training and a test set (70:30).

The acoustic and prosodic features of each audio recording were extracted using the \textsc{openSMILE-v2.4.1} toolkit~\citep{eyben2010opensmile}. Two different feature sets were extracted: 1. The extended Geneva Minimalistic Acoustic Parameter Set (\textsc{eGeMAPS})~\citep{eyben2015geneva}. This parameter set consists of 88 acoustic signal descriptors including 18 low-level descriptors  (e.\,g., pitch, jitter, formants, shimmer, loudness) to which different functionals were applied (e.\,g., arithmetic mean, standard deviation, percentiles), 6 temporal features (e.\,g., mean length of voiced and unvoiced regions), and 7 cepstral parameters (e.\,g., MFCC, spectral flux); 2. The Computational Paralinguistics Challenge (\textsc{ComParE\_2016})~\citep{schuller2016interspeech} parameter set, which contains 6,373 features. 

In the first step, the performances of four different classifiers -- Logistic Regression (LR), eXtreme Gradient Boosting (XGBoost), Support Vector Machine (SVM), and Random Forest (RF) -- on the two feature sets were measured. For each feature set and each of these classifiers, 21 different models were trained. The first ten models were trained on one language only and then tested on each of the ten languages separately (\textbf{single condition}). These models were also tested on a combined test set containing all languages except the one on which the model was trained (\textbf{multi-test condition}). The next ten models were trained on all languages except one and then tested on the remaining language (\textbf{multi-train condition}). The last model was trained and tested on all languages (\textbf{all condition}). This process was repeated five times and the mean performances were used for evaluation. 

Further hyperparameter optimisation did not yield any additional performance gains. Therefore, we only report results for the default parameters (e.g., for RF: n\_estimators = 100, max\_depth = None, min\_samples\_split = 2). For LR, SVM, and RF we used \textsc{sklearn-v1.2.0} and for XGBoost \textsc{xgboost-v1.7.3} in \textsc{Python-v3.10.06}.

Model performance was determined using the unweighted average recall (UAR) score, which compensates for imbalances in the sample class ratio by summing up the recall of all classes and dividing this by the number of classes, and the F1 score. The means of the multi-test and multi-train models were used to compare the classifiers. The best-performing feature set and classifier were selected for subsequent analysis.

These subsequent analyses included examining the most influential parameters for the model performance using SHapley Additive exPlanations (\textsc{SHAP}). In addition, Mann-Whitney \textit{U} (MWU) tests were used to examine differences in the value distributions between abusive and non-abusive recordings for each feature in each language. A difference was called meaningful if two conditions were met: 1. The MWU test result had to be significant using an alpha level of 0.05 and correcting for multiple testing within the language using the Bonferroni-Holm method; 2. The Common-Language Effect Size (CLES), which indicates ``the probability that a score sampled at random from the first population will be greater than a score sampled at random from the second''~\citep[p.~101]{Vargha2000CLES}, had to be greater than 67.2\%, which corresponds to a medium effect size. If a feature was meaningful in every language, it was considered important for the differentiation between abusive and non-abusive audio recordings. The findings from the MWU tests were compared with the \textsc{SHAP} results. The comparison as well as the interpretation of the results can be found in \cref{sec:discussion}.

\section{Results}
\label{sec:results}
\begin{figure*}[t]
    \includegraphics[width=\textwidth]{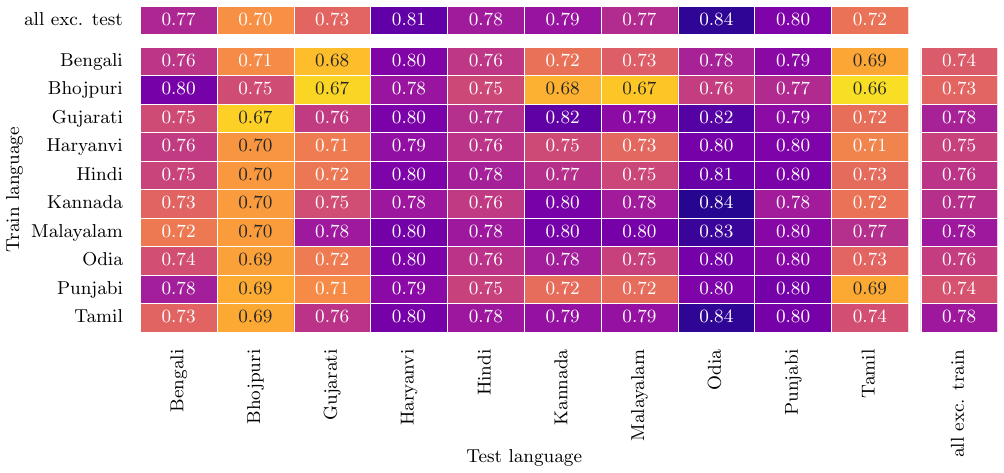}
    \caption{Heatmap of UAR scores for the RF classifier trained on each of the ten languages in the ADIMA data set using the \textsc{eGeMAPS} features. The rows indicate the training language, while the columns indicate the test language. The top row shows the performance when training on all languages except the test language; the rightmost column shows the performance when testing on all languages except the training language.}
    \label{figure:heatmap}
\end{figure*}

\begin{figure}[t]
    \includegraphics[width=\columnwidth]{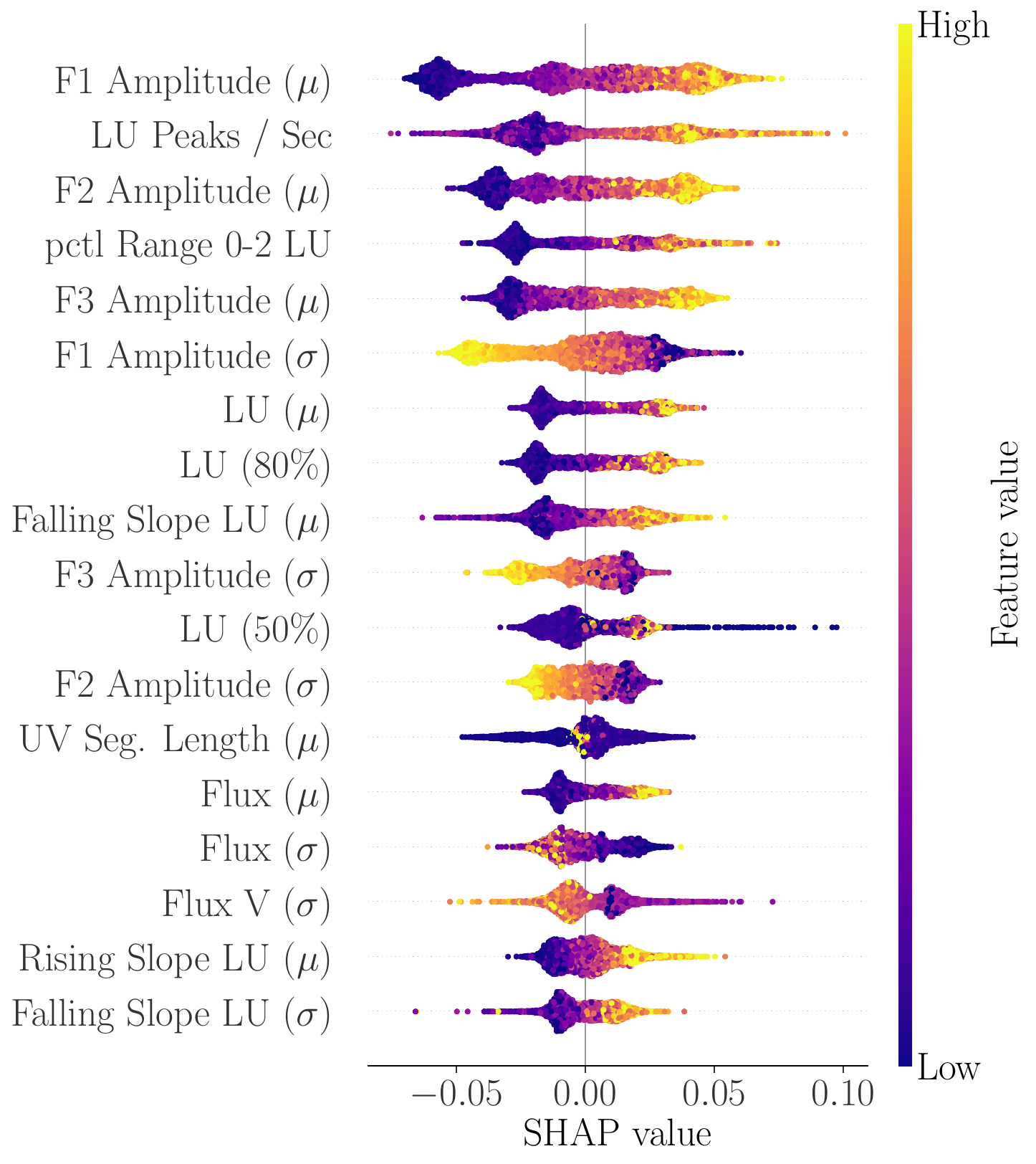}
    \caption{\textsc{SHAP} graph for the RF classifier trained on all ten languages in the ADIMA data set using the \textsc{eGeMAPS} features. The values indicate the influence of the features on the model. Note: LU = Loudness}
    \label{figure:shap}
\end{figure}

The mean of the UAR and F1 scores for the two feature sets and four classifiers can be found in \cref{table:classifiers}. In general, it can be said that the classifiers performed similarly well on the \textsc{eGeMAPS} feature set and on the \textsc{ComParE\_2016} feature set. Due to its size advantage, it was decided to use the \textsc{eGeMAPS} feature set for all subsequent analyses in this paper. As for the classifiers, all models seem to perform comparably well in the \textit{multi-train condition}, while in the \textit{multi-test condition}, the RF classifier performed slightly better than the others. For this reason, the RF classifier was used to determine all other results. The result for the \textit{all condition} can also be found in \cref{table:classifiers}, while all other results for the RF classifier, including multilingual and cross-lingual settings, can be found in \cref{figure:heatmap}. To investigate which features are most influential in the classification a \textsc{SHAP} graph was created using a model trained on all languages (see \cref{figure:shap}).

\begin{table}[t]
    \centering
    \caption{For both feature sets, the means of the UAR and F1 scores are shown for all classifiers using the models trained on all but one language and tested on the last (multi-train condition) and the models trained on one language and tested on all others (multi-test condition). In addition, the score for training and testing on all languages (all condition) is shown for the RF classifier.}
    \begin{tabular}{llrrrr}
        \toprule
        & & \multicolumn{2}{c}{\textsc{eGeMAPS}} & \multicolumn{2}{c}{\textsc{ComParE}} \\
        \cline{3-6}
        Classifier & Condition & UAR & F1 & UAR & F1 \\
        \midrule
        LR & multi-train & 0.77 & 0.72 & 0.68 & 0.63 \\
        LR & multi-test & 0.74 & 0.67 & 0.72 & 0.66 \\
        XGBoost & multi-train & 0.76 & 0.72 & 0.77 & 0.74 \\
        XGBoost & multi-test & 0.75 & 0.67 & 0.75 & 0.68 \\
        SVM & multi-train & 0.77 & 0.73 & 0.67 & 0.62 \\
        SVM & multi-test & 0.74 & 0.67 & 0.71 & 0.65 \\
        RF & multi-train & 0.77 & 0.73 & 0.77 & 0.73 \\
        RF & multi-test & 0.76 & 0.68 & 0.76 & 0.68 \\
        \midrule
        RF & all & 0.79 & 0.76 & 0.80 & 0.76 \\
        \bottomrule
    \end{tabular}
    \label{table:classifiers}
\end{table}

Between 19 and 25 features were found to be meaningful in each language as a result of the MWU tests. A total of 18 features were considered important, i.\,e., meaningful for all ten languages -- Loudness ($\mu$), Loudness (50\%), Loudness (80\%), Loudness (pctl Range 0-2), Rising Slope Loudness ($\mu$), Rising Slope Loudness ($\sigma$), Falling Slope Loudness ($\mu$), Falling Slope Loudness ($\sigma$), F1 Amplitude ($\mu$), F2 Amplitude ($\mu$), F3 Amplitude ($\mu$), Flux ($\mu$), Flux V ($\mu$), Flux UV ($\mu$), Loudness (Peaks Per Sec), Voiced Segments Per Second, Voiced Segment Length Per Sec ($\mu$), $\bar{E}_{rms}$. Since many of these are highly correlated, only a selection of the results is shown. \cref{table:importantfeatures} contains the mean, median, and the $1^{st}$ and $3^{rd}$ quartiles for these features, calculated on the whole data set and split by the label. It also shows the mean of the CLES for the MWU tests across all languages.

\begin{table*}[t]
  \centering
  \caption{Selection of features, that differed significantly between abusive and non-abusive content in all languages according to MWU test (Bonferroni-Holms corrected p-value \textless 0.05 and CLES \textgreater 67.2\%), with respective means, $1^{st}$ quartile, $2^{nd}$ quartile (median), $3^{rd}$ quartile, and mean CLES.}
  \begin{tabular}{lrrrrrrrrr}
    \toprule
    & \multicolumn{4}{c}{non-abusive} & \multicolumn{4}{c}{abusive} &  \\
    \cline{2-9}
    Features & \textit{M} & $Q_1$ & Median & $Q_3$ & \textit{M} & $Q_1$ & Median & $Q_3$ & CLES \\
    \midrule
    Loudness ($\mu$) & 0.41 & 0.07 & 0.23 & 0.54 & 1.37 & 0.50 & 1.00 & 2.03 & 0.82 \\
    F1 Amplitude ($\mu$) & -161.61 & -193.20 & -168.86 & -139.51 &  -109.48 & -141.90 & -110.56 & -76.12 & 0.82 \\
    F2 Amplitude ($\mu$) & -157.40 & -190.91 & -163.91 & -133.59 & -103.73 & -136.40 & -103.50 & -69.98 & 0.82 \\
    F3 Amplitude ($\mu$) & -158.50 & -190.81 & -164.67 & -135.82 & -106.34 & -138.47 & -105.90 & -73.20 & 0.82 \\
    Flux ($\mu$) & 0.25 & 0.02 & 0.12 & 0.31 & 0.94 & 0.28 & 0.63 & 1.41 & 0.82 \\
    Voiced Segments per Second & 1.16 & 0.40 & 1.00 & 1.72 & 2.13 & 1.37 & 2.08 & 2.80 & 0.76 \\
    \bottomrule
  \end{tabular}
  \label{table:importantfeatures}
\end{table*}

\section{Discussion}
\label{sec:discussion}
In this paper, we tried to investigate whether paralinguistic, i.\,e., acoustic and prosodic, features alone can be used to detect abusive content in real-life audio recordings. We also wanted to find out how these features perform in real multilingual and cross-lingual settings. The results show that it is indeed possible to use paralinguistic features to classify abusive and non-abusive content. For the RF classifier, the achieved UAR values were between 0.70 and 0.84 for multilingual settings and between 0.66 and 0.84 for cross-lingual settings. These results are comparable to those of related studies.

Comparing the \textsc{eGeMAPS} and \textsc{ComParE\_2016} feature sets, the results indicate that the \textsc{eGeMAPS} feature set performs comparably well when used for abusive content detection in audio data. In addition, it has the advantage of better interpretability and shorter run time due to its smaller size. Regarding the classifiers, the RF classifier performs best on the given data set and using the extracted paralinguistic features. This finding is consistent with a study by \citet{ibanez2021audio}, which attempted to detect hate speech in audio data, and in contrast to a study by \citet{soni2018see}, which found that LR performed best in detecting cyberbullying in audio and visual data. One explanation is the difference in the used features, as RF can model non-linear dependencies, unlike LR. Another problem may be that the features in the \textsc{eGeMAPS} feature set are highly correlated, which is a problem in LR and can lead to unreliable coefficient estimates.

As shown in \cref{figure:heatmap}, the multilingual models performed best when tested on Odia, Haryanvi, and Punjabi and worst on Bhojpuri. This was also true for most cross-lingual models. These results were present in all classifiers and can be found similarly in~\citep{gupta2022adima}. \citet{sharon2022multilingual}, on the other hand, obtained different results, with Bhojpuri performing very well and Odia performing worst. This difference may be due to the different properties of the languages, as \citet{ozseven2018investigation} shows that there are differences in the predictive properties between different languages.

The results of the MWU tests and the most influential features of the model, as seen in the \textsc{SHAP} graph, show distinct similarities. Both indicate that the most important features in determining whether a recording is abusive or non-abusive are the mean amplitudes of F1, F2, and F3, the loudness features, and the mean spectral flux -- with higher values indicating abusive language. This is reminiscent of results from the anger literature, which found higher mean volume, higher volume range, and higher energy density \citep{sobin1999emotion}, as well as higher F1, F2, and F3 \citep{yuan2002acoustic, mohanta2016classifying, ozseven2018investigation}, as significant predictors. In addition, the \textsc{SHAP} plot suggests that abusive content is associated with lower standard deviation in F1, F2, and F3 amplitudes. In contrast to other studies, we did not find a difference in pitch~\citep{liscombe2003classifying, ozseven2018investigation}. This is consistent with preliminary evidence for lower or absent pitch variation in (acted) emotional speech in Indian languages~\citep{Rao13-ERS, Mathew10-AES}; as research in these languages is still in its infancy~\citep{Ignatius21-SPT} though, these findings should be revisited. 

In general, the features found to be most influential in our study seem to be mostly related to angry speech. Other mechanisms used to convey abusive content, such as sarcasm and irony~\citep{poletto2021resources}, are either not covered in the ADIMA data set or are not captured by our features. This alternative form of abusive speech may be more difficult to detect using prosody alone~\citep{bryant2005there}. In addition, \citet{dougdu2022comparison} have shown that \textsc{eGeMAPS} has a tendency to lead to confusion of happiness and anger categories in different classifiers, which may be due to the similarity in arousal between these two categories~\citep{kuchibhotla2014comparative}. This raises the question of whether our classifiers also capture arousal. 

One problem with the current study that needs to be addressed is the lack of age and sex information in the data set. Therefore, it is not possible to determine whether the data set is balanced with respect to these attributes. Since acoustic features typically change with age and also differ between males and females, this could explain some differences in features, especially in pitch and vowel formant frequency, which are typically lower for males, and voice onset time. In addition, \citet{mohanta2016classifying} showed that the anger distinguishing features differ between males (pitch as the main predictor) and females (F1, F2, and F3 as main predictors).

\section{Conclusions}
\label{sec:conclusion}
The need to identify abusive content in audio and video data is growing rapidly. This study shows that paralinguistic features, i.\,e., acoustic and prosodic properties, can be successfully used to classify real-life audio recordings depending on the presence of abusive and non-abusive content. Using the features from \textsc{eGeMAPS}, we were able to achieve UAR scores between 0.66 and 0.84 for cross-lingual settings. The multilingual performance is particularly interesting, showing that a model trained on all languages except one still performs well on the excluded language. 

The most influential features in our model were the mean loudness, mean F1, F2, and F3 amplitude, mean flux, and mean voiced segments per second. These results suggest that angry speech played a critical role in detecting abusive content. Future research should further investigate how to detect other mechanisms such as sarcasm and irony.

\textbf{Breaking Down Barriers:} Concerning this year's conference theme, we believe the detection of abusive speech to be a highly relevant problem in the current internet and social media era. Moreover, ADIMA is a multi-lingual data set from a non-English domain and thus contributes to inclusive research in speech science by targeting a diverse audience. However, it does not readily contain metadata such as age and sex; we were, therefore, unable to evaluate the (intersectional) generalisation of our methods with respect to these variables.

\section{Acknowledgements}
This work has received funding from the EU’s Horizon 2021 grant agreement No.\ 101060660 (SHIFT).

\section{\refname}
\printbibliography[heading=none]

\end{document}